\documentclass[12pt]{iopart}

\usepackage{iopams}
\usepackage{graphicx}
\usepackage{subfigure}
\usepackage{textcomp}
\usepackage{bm}

\begin{document}

\title[Magnetic field imaging with NV ensembles]{Magnetic field imaging with NV ensembles}

\author{L. M. Pham}
  \address{School of Engineering and Applied Sciences, Harvard University, Cambridge, MA 02138, USA}
\author{D. Le Sage}
  \address{Harvard-Smithsonian Center for Astrophysics, Cambridge, MA 02138, USA}
\author{P. L. Stanwix}
  \address{Harvard-Smithsonian Center for Astrophysics, Cambridge, MA 02138, USA}
\author{T. K. Yeung}
  \address{Physics Department, Harvard University, Cambridge, MA 02138, USA}
  \address{School of Engineering and Applied Sciences, Harvard University, Cambridge, MA 02138, USA}
\author{D. Glenn}
  \address{Harvard-Smithsonian Center for Astrophysics, Cambridge, MA 02138, USA}
\author{A. Trifonov}
  \address{Physics Department, Harvard University, Cambridge, MA 02138, USA}
  \address{Ioffe Physical-Technical Institute RAS, Saint Petersburg, Russia}
\author{P. Cappellaro}
  \address{Nuclear Science and Engineering Department, Massachusetts Institute of Technology, Cambridge, Massachusetts 02139, USA}
\author{P. R. Hemmer}
  \address{Electrical and Computer Engineering Department, Texas AM University, College Station, TX, 77843, USA}
\author{M. D. Lukin}
  \address{Physics Department, Harvard University, Cambridge, MA 02138, USA}
\author{H. Park}
  \address{Physics Department, Harvard University, Cambridge, MA 02138, USA}
  \address{Chemistry and Chemical Biology Department, Harvard University, Cambridge, MA 02138, USA}
\author{A. Yacoby}
  \address{Physics Department, Harvard University, Cambridge, MA 02138, USA}
\author{R. L. Walsworth}
  \address{Harvard-Smithsonian Center for Astrophysics, Cambridge, MA 02138, USA}
  \address{Physics Department, Harvard University, Cambridge, MA 02138, USA}
  \ead{rwalsworth@cfa.harvard.edu}

\begin{abstract}
We demonstrate a method of imaging spatially varying magnetic fields using a thin layer of nitrogen-vacancy (NV) centers at the surface of a diamond chip. Fluorescence emitted by the two-dimensional NV ensemble is detected by a CCD array, from which a vector magnetic field pattern is reconstructed. As a demonstration, AC current is passed through wires placed on the diamond chip surface, and the resulting AC magnetic field patterns are imaged using an echo-based technique with sub-micron resolution over a 140 $\rm{\mu m}$ $\times$ 140 $\rm{\mu m}$ field of view, giving single-pixel sensitivity $\rm{\sim 100\:nT/\sqrt{Hz}}$. We discuss ongoing efforts to further improve sensitivity and potential bioimaging applications such as real-time imaging of activity in functional, cultured networks of neurons.
\end{abstract}

\maketitle

\section{Introduction}
The negatively charged nitrogen-vacancy (NV) color center in diamond exhibits several remarkable properties, which make it a promising system for applications in quantum information processing~\cite{ChildressSci2006, WrachtrupJPhys2006} and magnetic field sensing~\cite{TaylorNatPhys2008, MazeNat2008, BalasubramanianNat2008, DegenAPL2008}. Chief among these properties are the NV center's electronic structure, which allows for optical initialization and detection of the spin state, and long room temperature spin coherence times. These long coherence times are of particular relevance to magnetic sensing applications. For example, in high purity samples with natural 1.1\% abundance of $\rm{^{13}C}$, single NV centers with $\rm{\gtrsim 600}$ $\rm{\mu s}$ coherence times have demonstrated 30 $\rm{nT/\sqrt{Hz}}$ magnetic field sensitivity~\cite{MazeNat2008}, while single centers in 0.3\% $\rm{^{13}C}$ isotopically engineered samples have achieved coherence times $\rm{\approx}$ 2 ms and magnetic field sensitivity $\rm{\approx 4}$ $\rm{nT/\sqrt{Hz}}$~\cite{BalasubramanianNatMat2009}. A natural extension of such results is to demonstrate magnetometry using spatially extended ensembles of NV centers, resulting in a larger fluorescence signal and the ability to image magnetic field patterns. In this paper, we discuss a method for 2D imaging of magnetic fields using an ensemble of NV centers in a thin layer at the surface of a diamond chip. We present preliminary data for this NV ensemble magnetic field imager and discuss routes to optimization and application to imaging neural network dynamics.

\begin{figure}
\subfigure[]{
\includegraphics[width=0.24\columnwidth]{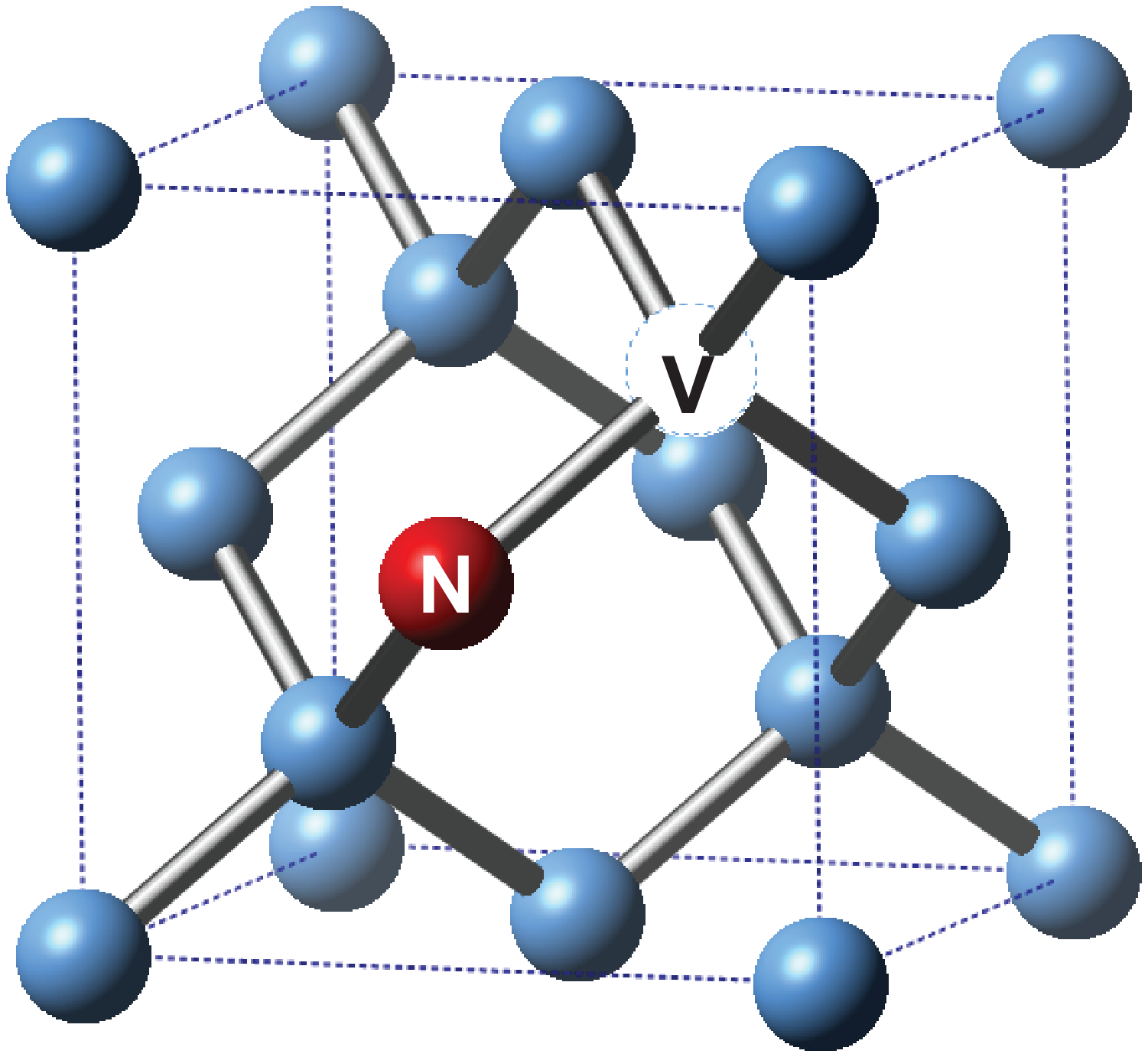}
\label{fig:background:physical}
}
\subfigure[]{
\includegraphics[width=0.31\columnwidth]{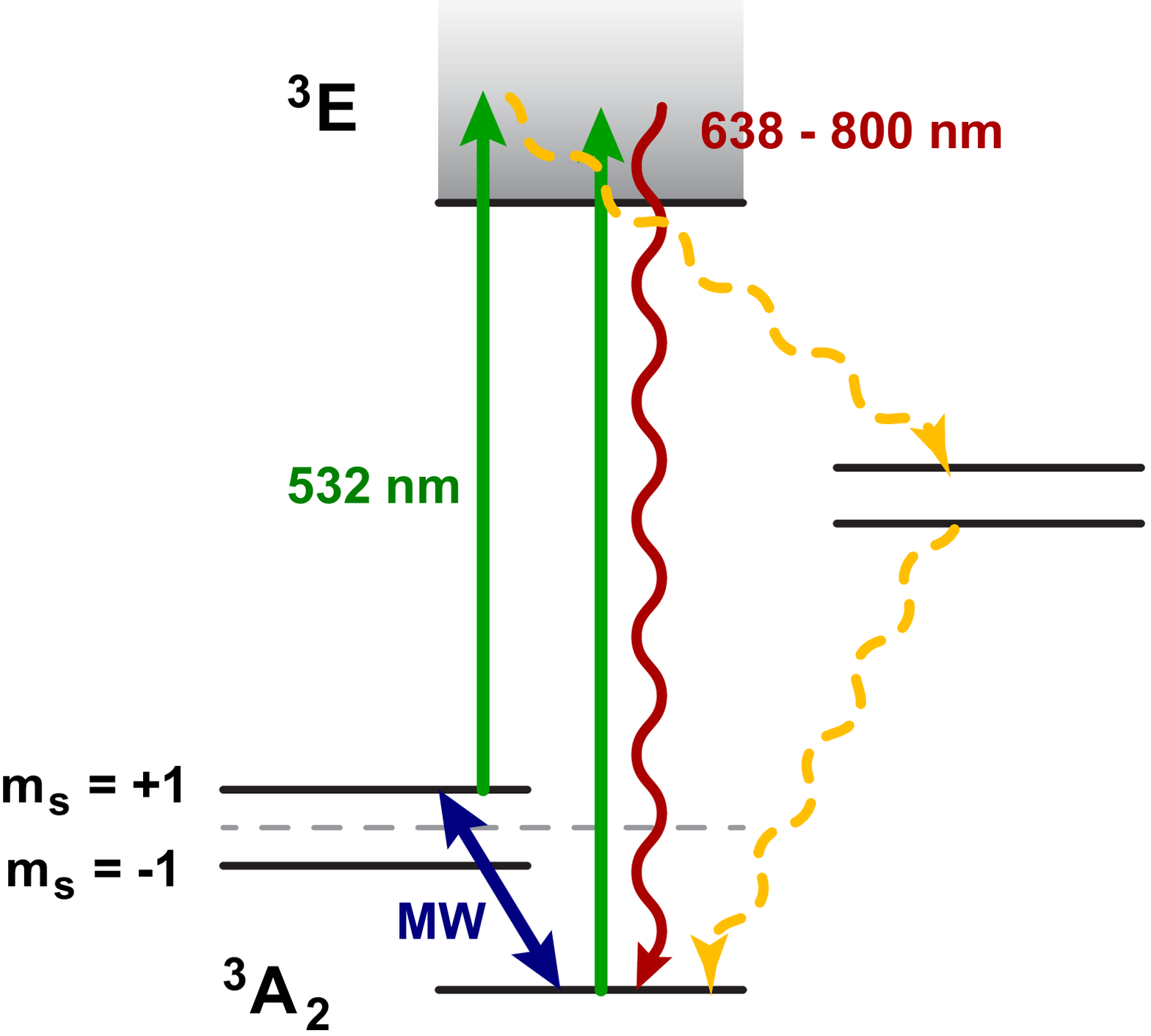}
\label{fig:background:electronic}
}
\subfigure[]{
\includegraphics[width=0.39\columnwidth]{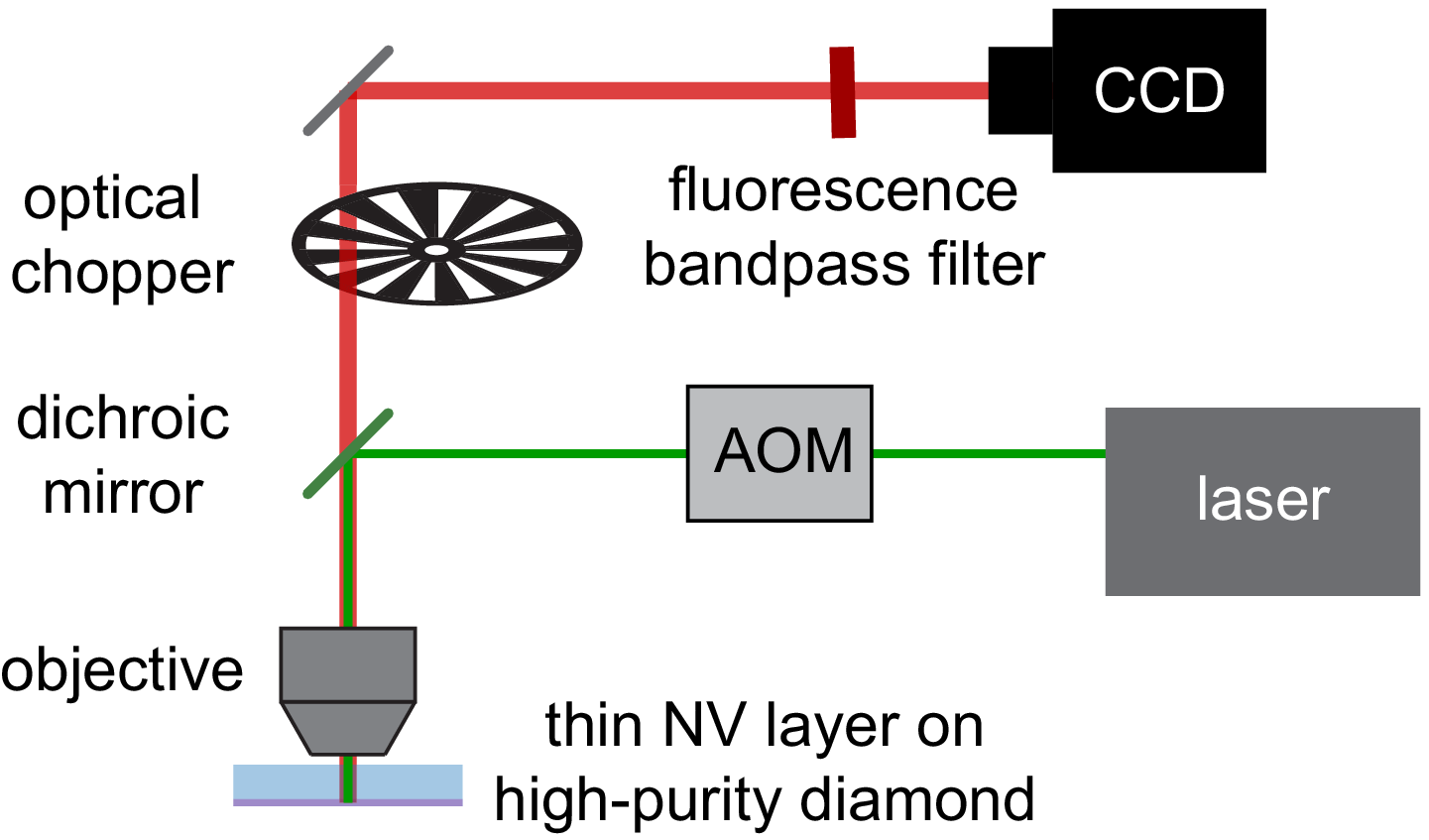}
\label{fig:setup:schematic}
}
\subfigure[]{
\includegraphics[width=0.50\columnwidth]{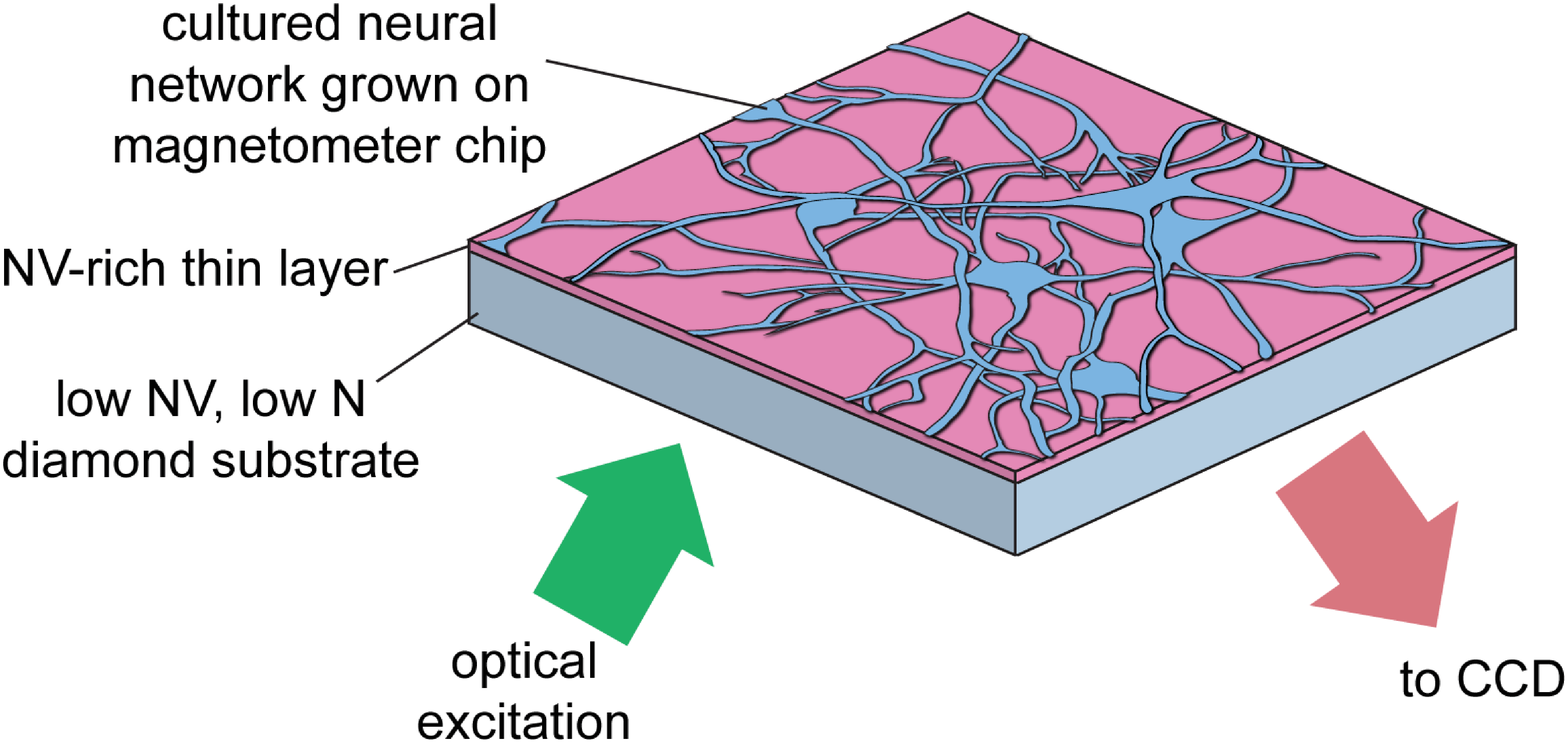}
\label{fig:background:neurodiagram}
}
\subfigure[]{
\includegraphics[width=0.43\columnwidth]{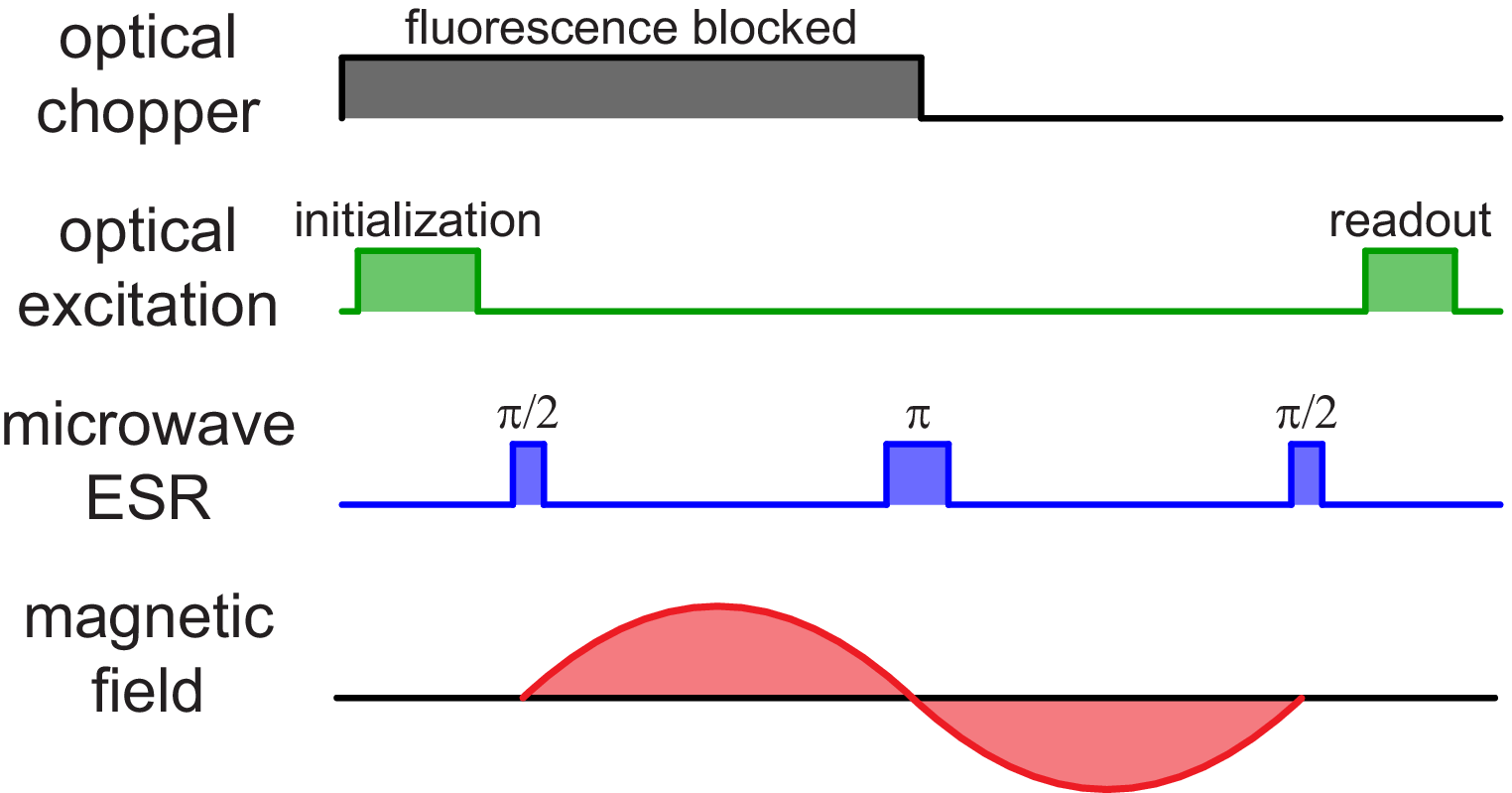}
\label{fig:setup:pulseseq}
}
\label{fig:background}
\caption{~\subref{fig:background:physical} The NV electronic spin axis is defined by nitrogen and vacancy sites, in one of four crystallographic directions. NV orientation subsets in an ensemble can be spectrally selected by applying a static magnetic field.~\subref{fig:background:electronic} NV center electronic energy level structure. Spin polarization and readout is performed by optical excitation and fluorescence detection. Ground state spin manipulation is achieved by resonant microwave excitation. The ground state triplet has a zero magnetic field splitting $\rm{\simeq}$ 2.87 GHz.~\subref{fig:setup:schematic} Schematic of the NV ensemble magnetic field imager. ~\subref{fig:background:neurodiagram} Illustration of the NV ensemble magnetic field imager and its potential application in mapping neuronal network dynamics.~\subref{fig:setup:pulseseq} Hahn-echo pulse sequence used to perform AC magnetometry.}
\end{figure}

The NV center is composed of a substitutional nitrogen atom (N) and a vacancy on adjacent lattice sites (Fig.~\ref{fig:background:physical}). The negatively charged state, which is the focus of this work, is believed to gain an electron from nearby N donor impurities~\cite{GaebelAPB2006}. The energy-level diagram of an NV center is depicted in Fig.~\ref{fig:background:electronic}. The NV center has a spin-triplet ground state with a zero-field splitting of $\rm{\sim}$ 2.87 GHz between the $\rm{m_s = 0}$ and $\rm{m_s = \pm1}$ spin states, quantized along the nitrogen-vacancy axis. Applying a small external magnetic field along this axis lifts the degeneracy of the $\rm{m_s = \pm1}$ energy levels with a Zeeman shift of $\rm{\sim}$ 2.8 MHz/G. Optical transitions between the ground and excited states have a characteristic zero-phonon line (ZPL) at 637 nm with a broad phonon-sideband at room temperature. These transitions are primarily spin conserving; however, there exists an alternative decay path that transfers approximately 30\% of the $\rm{m_s = \pm1}$ excited state population to the $\rm{m_s = 0}$ ground-state through a metastable singlet state~\cite{MansonPRB2006, AcostaPRB2010}, without emitting a photon in the fluorescence band. These key characteristics of the NV center allow its electronic spin-state to be prepared through optical pumping and coherent ground-state manipulation, and read out via spin-state dependent fluorescence.

In this work, we consider the application of the NV center and its electronic spin to sensitive detection of AC magnetic fields. First, an initial superposition of NV spin states is formed; then, the NV spins accumulate relative phase in the presence of an external AC magnetic field during a field-interrogation period. By mapping the resulting state into spin populations, which can be measured optically, information about the external field can be extracted~\cite{TaylorNatPhys2008}. The optimal photon shot noise limited sensitivity of such a magnetometer is given in terms of the minimum detectable variation in AC magnetic field amplitude $B$ and the total measurement time $T$ by the following formula:
\begin{equation}\label{eq:sens}
\eta \equiv \delta B \sqrt{T} \approx \frac{\pi \hbar}{2 g \mu_{B}} \frac{1}{C \sqrt{N \tau}}
\end{equation}
where $C$ is a parameter which encompasses the signal contrast and number of photons detected per NV, $N$ is the number of NVs probed, and $\tau$ is the field-interrogation period, optimized at the $\rm{T_2}$ coherence time for AC magnetic field measurements~\cite{TaylorNatPhys2008}.

The sensitivity of an NV-based magnetic field sensor can thus be improved by increasing the interrogation time through a longer $\rm{T_2}$ coherence time and by increasing the signal through improved collection efficiency or probing more NVs. There has been recent work toward improving the sensitivity of static magnetic field measurements using ensembles instead of single NV centers, with application to vector magnetometry and magnetic field imaging~\cite{MaertzAPL2010, SteinertRevSciIns2010, AcostaAPL2010, SchoenfeldPRL2011}. However, the sensitivity of such DC magnetometry schemes is limited by typical $\rm{\mu s}$-scale NV $\rm{T_2^*}$ coherence times. Here we apply an echo-based AC magnetometry measurement technique to ensembles of NV centers, to take advantage of the much longer $\rm{T_2}$ coherence time ($\sim 100$ $\rm{\mu s}$) and the increased signal provided by an NV ensemble.

Furthermore, the extension of NV magnetometry to ensembles lends itself to a CCD-based rapid imaging scheme that is not possible when probing a single center. By exciting NV centers confined to a thin surface layer over a large area in bulk diamond and imaging the resulting fluorescence onto a CCD camera, we demonstrate optical detection of 2D AC magnetic field patterns over large fields of view ($>100$ $\rm{\mu m}$) with sub-micron spatial resolution and sensitivity $\sim 100$ $\rm{nT/\sqrt{Hz}}$ per pixel.

\section{NV Ensemble Coherence}
In order to take full advantage of the dual improvements in sensitivity afforded by performing AC measurements and probing an ensemble of NV centers, the ensemble must exhibit a long $\rm{T_2}$ coherence time. This time is limited by spin impurities in the diamond, such as $\rm{^{13}C}$ and N atoms. Recent NV ensemble measurements of $T_2$ in diamond samples with low nitrogen content and a natural abundance of $\rm{^{13}C}$ isotopes demonstrated the same long $\rm{\gtrsim}$ 600 $\rm{\mu s}$ $\rm{T_2}$ coherence times as single centers in comparable samples~\cite{StanwixPRB2010}. In such samples, where $\rm{T_2}$ is limited by the $\rm{^{13}C}$ bath, the measured ensemble coherence times are strongly dependent on small misalignments between the static magnetic field and the NV axis.

\begin{figure}
  \includegraphics[width=0.85\columnwidth]{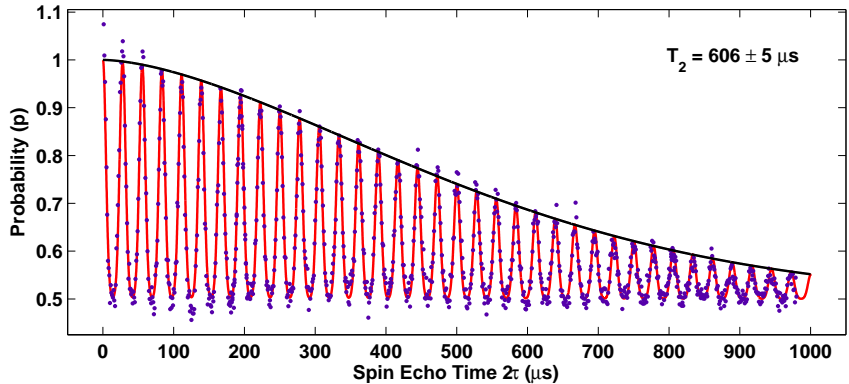}
  \label{fig:13CT2:longT2}
\caption{Long coherence time of 606 $\rm{\mu s}$ measured from an ensemble of NV centers in a CVD diamond sample with a $\rm{\approx}$ 1 $\rm{\mu m}$ thin layer of NV density $\rm{\approx}$ 3 $\times$ $10^{13}$ $\rm{cm^{-3}}$ and N density $\rm{\approx}$ 1 $\times$ $10^{15}$ $\rm{cm^{-3}}$ on a $\rm{\approx}$ 200 $\rm{\mu m}$ diamond substrate of very low NV and N concentration. This long ensemble NV $\rm{T_2}$ is comparable to coherence times measured from single NV centers in similar low N impurity samples.}
\end{figure}

In the presence of an applied static magnetic field, each NV center experiences fluctuating dipolar magnetic fields due to the precession of $\rm{^{13}C}$ nuclei. When the static magnetic field is aligned with the NV symmetry axis, the field from each $\rm{^{13}C}$ nucleus averages to zero over free precession intervals that are integer multiples of the $\rm{^{13}C}$ Larmor precession period, resulting in the characteristic collapses and revivals observed in NV Hahn-echo data for both single NV centers and NV ensembles (see Fig.~\ref{fig:13CT2:longT2}). However, when there is a small angle misalignment between the static magnetic field and the NV axis, the Larmor frequencies of $\rm{^{13}C}$ nuclear spins that are proximal to an NV center are altered by hyperfine interactions with the NV. The change in Larmor frequency for individual $\rm{^{13}C}$ nuclei is dependent on the component of the static magnetic field transverse to the NV axis and the particular location of each $\rm{^{13}C}$ nucleus relative to the NV center. Given a random distribution of 13C nuclei in the diamond lattice, the incommensurate $\rm{^{13}C}$ Larmor precession frequencies result in incomplete NV Hahn-echo revivals and an effectively shortened $\rm{T_2}$ that is strongly dependent on misalignment of the static magnetic field from the NV axis~\cite{StanwixPRB2010, MazePRB2008}.

Diamonds with a higher concentration of NV centers will produce a larger fluorescence signal, but in general also contain a higher concentration of nitrogen paramagnetic impurities, which further limits the $T_2$ coherence time. Although such samples have shorter coherence times, they are also less sensitive to small misalignments in the static field. In the development of a practical NV-based magnetic field imager, design parameters such as the required sensitivity, the alignment of the static field with the NV axis, and the range of AC field frequencies that can be measured will dictate the desired level of impurity concentration.

\section{NV Ensemble Magnetic Field Imager: Preliminary Results}
In order to demonstrate the imaging capabilities of NV ensembles in diamond, we custom built a wide-field fluorescence microscope (Fig.~\ref{fig:setup:schematic}). A 3-Watt 532 nm laser (Laser Quantum, Opus) optically excites a large region of interest on the diamond surface through a 20x 0.75NA objective (Nikon). An acousto-optic modulator (Isomet, M1133-aQ80L-H) acts as an optical switch, producing timed laser pulses to prepare and detect the NV spin states. A loop antenna is positioned near the diamond surface and connected to the amplified output of a microwave signal generator (Agilent, E8257D) to generate a homogeneous $\rm{B_1}$ field over the region of interest. Fast-switching of the microwave field allows for the coherent manipulation of the NV spin states necessary for magnetometry measurements.

The AC magnetometry measurement sequence is depicted in Fig.~\ref{fig:setup:pulseseq}. We use a standard Hahn-echo microwave pulse sequence to measure the external applied field~\cite{TaylorNatPhys2008, HahnPhysRev1950}. During the readout laser pulse, the fluorescence from the NVs is collected by the objective, filtered, and imaged onto a cooled CCD camera (Starlight Xpress, SXV-H9). Since the duration of a single magnetic field measurement is much shorter than the minimum exposure time of the camera, several thousand measurements are integrated within a single exposure. An optical chopper synchronized with the measurement sequence acts as a shutter for the camera during each spin-initialization laser pulse. Repeating the measurement without the microwave control pulses provides a reference for drifts in the laser intensity. Vector magnetometry is accomplished by resonantly coupling the microwaves to one of the four possible NV orientations and sequentially measuring the magnetic field projection onto the different NV vectors. Such a wide-field microscope geometry has the key advantages of being relatively simpler to construct and having a shorter intrinsic image acquisition time over a wide field of view than a scanning confocal microscope.

\subsection{Micron-Scale Magnetic Field Patterns}
Resolving magnetic field patterns which vary over sub-micron length scales places additional constraints on the diamond material properties beyond optimizing NV density and coherence time. Small field variations become irresolvable at distances greater than their feature size; therefore, in order to image magnetic fields on a sub-micron length scale, the NV sensors must be confined to a thin layer on the diamond surface and placed within a sub-micron distance from the magnetic field source. This thin-layer geometry can be achieved either by shallow implantation of nitrogen ions ~\cite{RabeauAPL2006, MeijerAPL2005} or by altering the growth conditions of chemical vapor deposition (CVD) synthesized diamonds such that there is a thin N-doped NV-rich layer at one surface of a high-purity substrate.

At the current state of diamond processing technology, shallow implantation offers finer control over the thickness of the nitrogen-doped layer than CVD growth does. Hence, to demonstrate the ability of the NV ensemble magnetic field imager to resolve field variations on the micron scale, we used a high-pressure, high-temperature (HPHT) high-purity synthesized diamond (Sumitomo) that had been implanted with $\rm{N_2^+}$ ions at 15 keV, creating a shallow NV layer $\sim$10 nm thick, $\sim$11 nm below the surface (as estimated from Monte-Carlo simulations using SRIM~\cite{ZieglerSRIM2008}). From the implant dose and the approximate N-to-NV conversion efficiency, we estimate that the 2D NV layer has an NV surface density of $\rm{\sim}$ 8 $\times$ $10^{10}$ $\rm{cm^{-2}}$ and an N density of $\rm{\sim}$ 1 $\times$ $10^{12}$ $\rm{cm^{-2}}$, resulting in a paramagnetic nitrogen-limited $\rm{T_2}$ of $\rm{\sim}$ 30 $\rm{\mu s}$.

\begin{figure}
\subfigure[]{
\includegraphics[width=0.48\columnwidth]{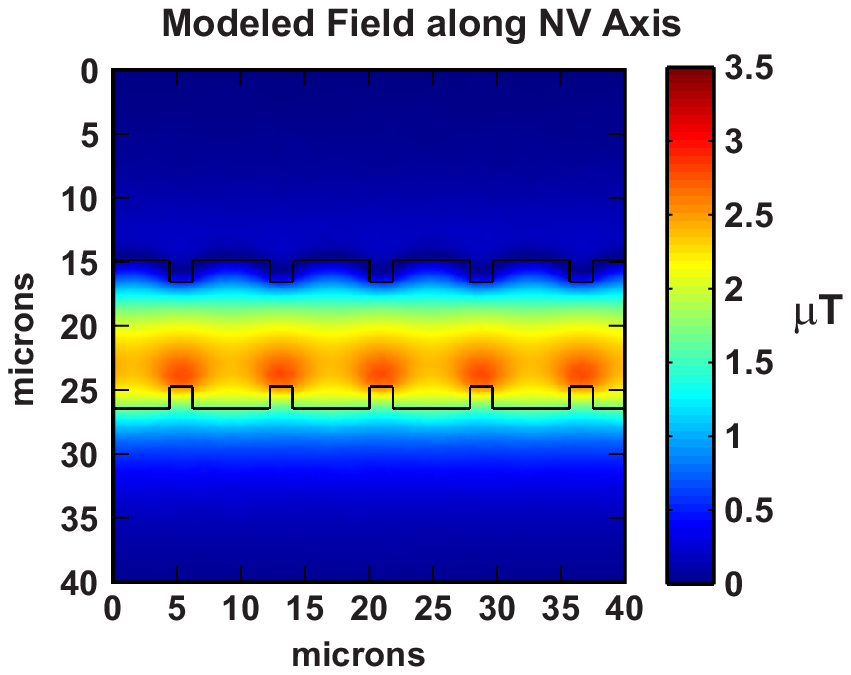}
\label{fig:notched:model}
}
\subfigure[]{
\includegraphics[width=0.48\columnwidth]{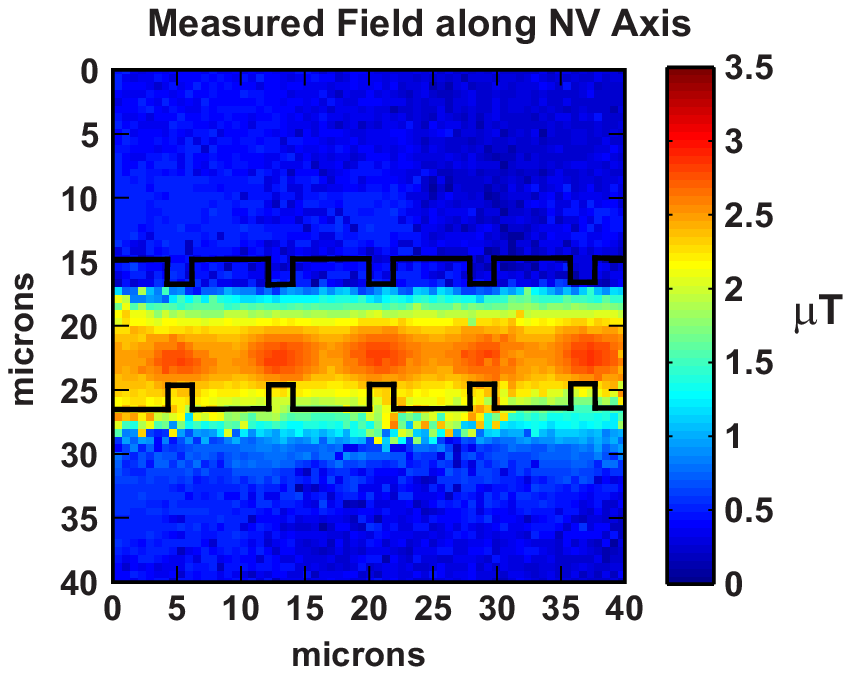}
\label{fig:notched:data}
}
\label{fig:notched}
\caption{~\subref{fig:notched:model} Modeled and~\subref{fig:notched:data} measured map of the field projection along the NV axis produced by running an AC current of frequency 109.5 kHz and amplitude 50 $\mu A$ through the notched wire pattern at a stand-off distance $\rm{\approx 1}$ $\rm{\mu m}$.}
\end{figure}

The magnetic field pattern to be imaged was produced by a copper microwire containing segments of alternating width such that the current density through the wire is increased in the constricted sections, resulting in micron-scale magnetic field `hotspots' (Fig.~\ref{fig:notched:model}). This notched wire pattern was fabricated on a glass coverslip using photolithography and secured flat against the shallow implant NV layer of the diamond sample with a separation distance of $\rm{\sim}$ 1 $\rm{\mu m}$. The proximity of the copper microwire to the NV sensors introduces a complication: when the microwave pulses are applied to coherently manipulate the NV spin states, the presence of the metal wire pattern modifies the microwave field profile, resulting in $\rm{B_1}$ field inhomogeneity over the plane of NV sensors. This microwave pickup is not expected to be an issue in the proposed bio-applications for the NV ensemble magnetic field imager, where the sources of the magnetic field are, for example, firing neurons rather than metal.

We used a permanent magnet to apply a static field along one of the NV axes perpendicular to the direction of current flow and ran 109.5 kHz AC current through the notched wire pattern. The fluorescence as a function of AC magnetic field amplitude was calibrated on a pixel-by-pixel basis by sweeping the AC current in order to account for spatial variations in the laser intensity and number of NV centers probed. In principle, such a calibration can also be accomplished using a homogeneous external AC field before measuring the field source of interest.

Applying microwave pulses resonant only with one NV orientation, we extracted from the NV fluorescence intensity an image of the AC magnetic field projection along the chosen NV axis (Fig.~\ref{fig:notched:data}). Furthermore, the measured magnetic field map agrees very closely with our modeled field map. Discrepancies can be attributed to the position and orientation of the wire pattern with respect to the crystallographic structure of the diamond sample being slightly different from the estimated values used in the model. We suspect that a small drift of the sample mount position over the course of the measurement is responsible for the slight excess blurring in the measured magnetic field image.

The sensitivity of the NV magnetometer (see Eq.~\ref{eq:sens}) can be measured experimentally as $\eta \equiv \delta B \sqrt{T} = \sigma_{m}\sqrt{T}/dS_{B}$, where $\sigma_{m}$ is the standard error in a set of fluorescence signal measurements, and $dS_{B}$ is the slope of the fluorescence signal as a function of the AC magnetic field amplitude~\cite{MazeNat2008}. To determine the sensitivity of the NV magnetic field microscope under the measurement conditions given above, we applied a uniform 109.5 kHz AC field across the region of interest and measured the standard deviation of the fluorescence signal at each pixel. The measurements were conducted at an applied AC field with maximum slope in order to optimize the sensitivity, and the average-best sensitivity was calculated as the mean single-pixel sensitivity in a 25 $\rm{\mu m}$ $\rm{\times}$ 25 $\rm{\mu m}$ area at the center of the gaussian laser excitation profile. Taking into account the time $T$ required to make a magnetic field measurement, the resulting average-best sensitivity per 614 nm $\times$ 614 nm pixel was $\rm{\approx}$ 9 $\rm{\mu T/ \sqrt{Hz}}$. Substituting experimental parameters such as mean photons collected per measurement, contrast, and measurement time, we calculated the expected shot-noise limited sensitivity per pixel to be $\rm{\approx}$ 6 $\rm{\mu T/\sqrt{Hz}}$. The measured  sensitivity is therefore in reasonable agreement with the shot-noise limited expected value~\cite{TaylorNatPhys2008}. We attribute the discrepancy to technical noise from the laser, which can be actively stabilized for future work.

\subsection{Wide-Field Magnetic Field Patterns}
The NV magnetic field microscope is also capable of measuring magnetic field variations over a wide field of view simultaneously. For this demonstration, the magnetic field pattern was generated by counter-propagating AC currents through a pair of copper wires in a zigzag pattern with $\rm{\sim}$ 100 $\rm{\mu m}$ features. The field was measured by an ensemble of NV sensors occupying a thin layer near the surface of a diamond sample held flat against the copper strips.

Since the magnetic field pattern in this demonstration does not exhibit large variations in field at a sub-micron level, it was not necessary to use the $\rm{N_2^+}$ ion-implanted diamond sample as above. Instead, we employed a CVD-grown sample (Apollo Diamond, Inc.) in which the bulk of the chip was grown in a low-nitrogen environment to produce a high-purity substrate. Additional nitrogen impurities were introduced near the end of the growth process, producing a $\rm{\sim}$ 3 $\rm{\mu m}$ thick layer of N-doped, NV-rich diamond at the surface of the otherwise pure substrate, which was confirmed by measuring the NV fluorescence as a function of depth using a scanning confocal microscope. From the growth conditions and the approximate N-to-NV conversion efficiency, we estimate that the NV layer has an NV density of $\rm{\sim}$ 1 $\times$ $10^{14}$ $\rm{cm^{-3}}$ and an N density of $\rm{\lesssim}$ 5 $\times$ $10^{15}$ $\rm{cm^{-3}}$, resulting in a paramagnetic nitrogen-limited $\rm{T_2}$ of $\rm{\sim}$ 280 $\rm{\mu s}$. The order of magnitude improvement in coherence time allows for more sensitive magnetometry; hence, such a CVD-grown sample is preferable in situations where the magnetic field varies on the scale of several microns.

\begin{figure}
\subfigure[]{
\includegraphics[width=0.5\columnwidth]{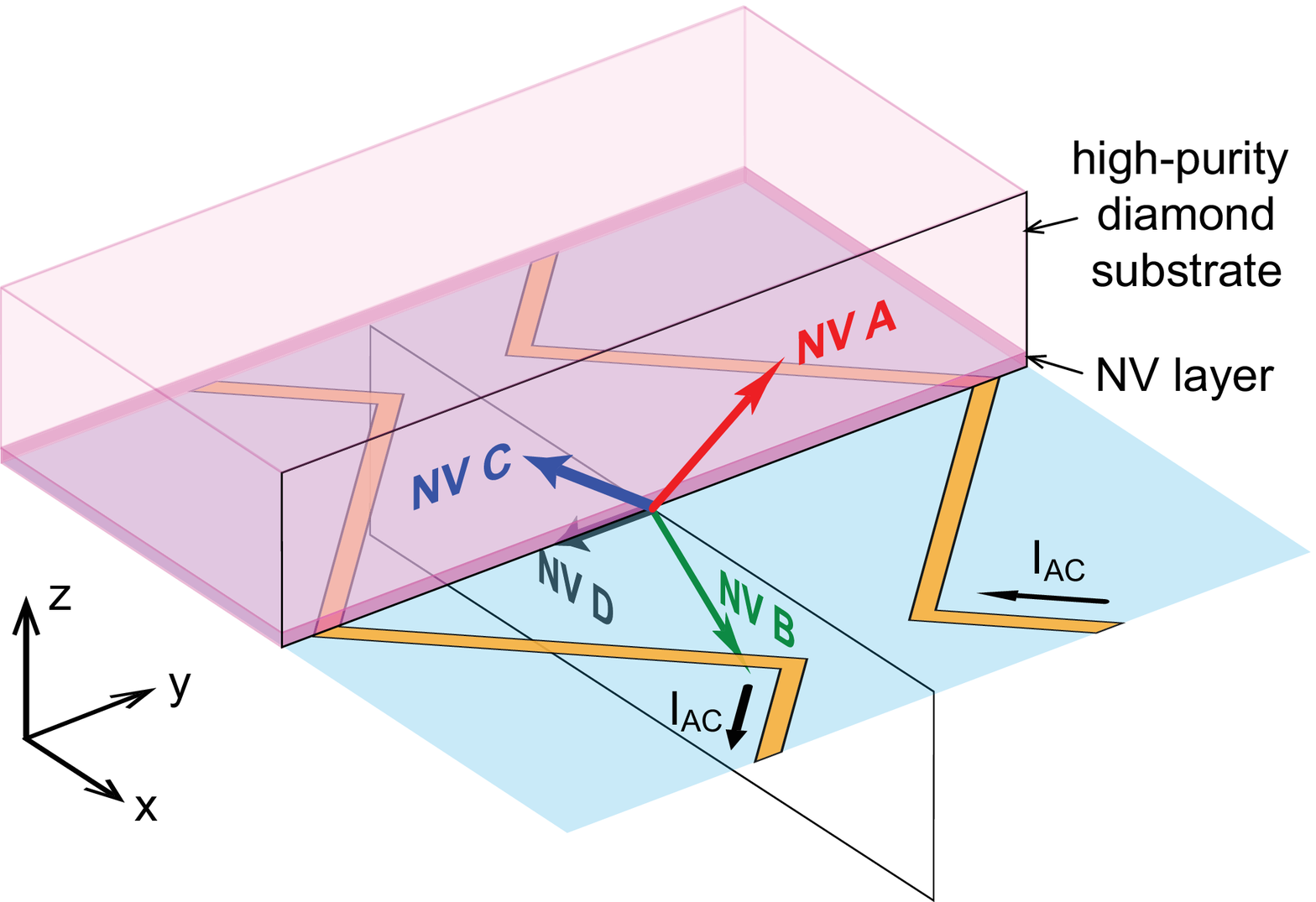}
\label{fig:zigzag:diagram}
}
\subfigure[]{
\includegraphics[width=0.4\columnwidth]{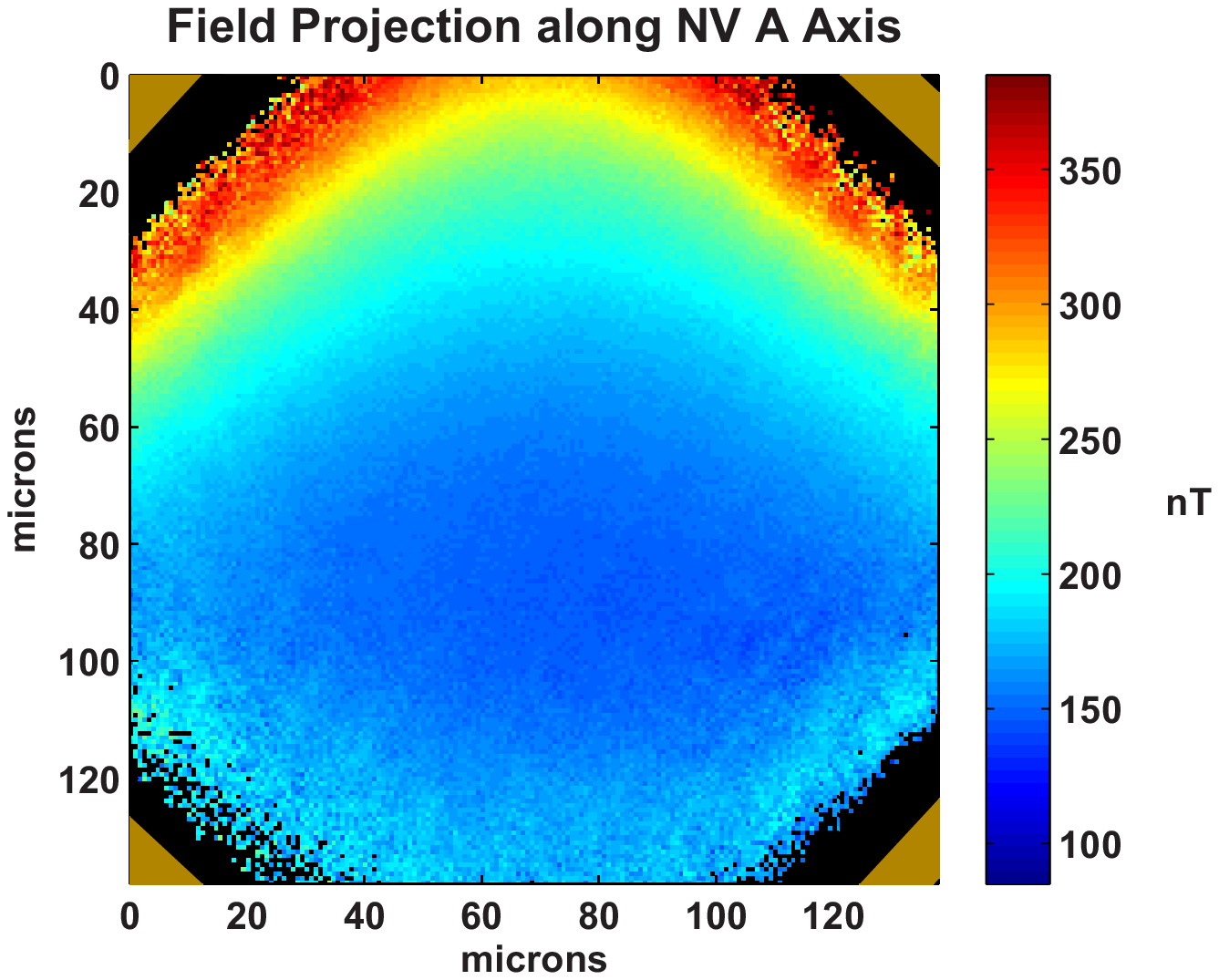}
\label{fig:zigzag:nva}
}
\subfigure[]{
\includegraphics[width=0.56\columnwidth]{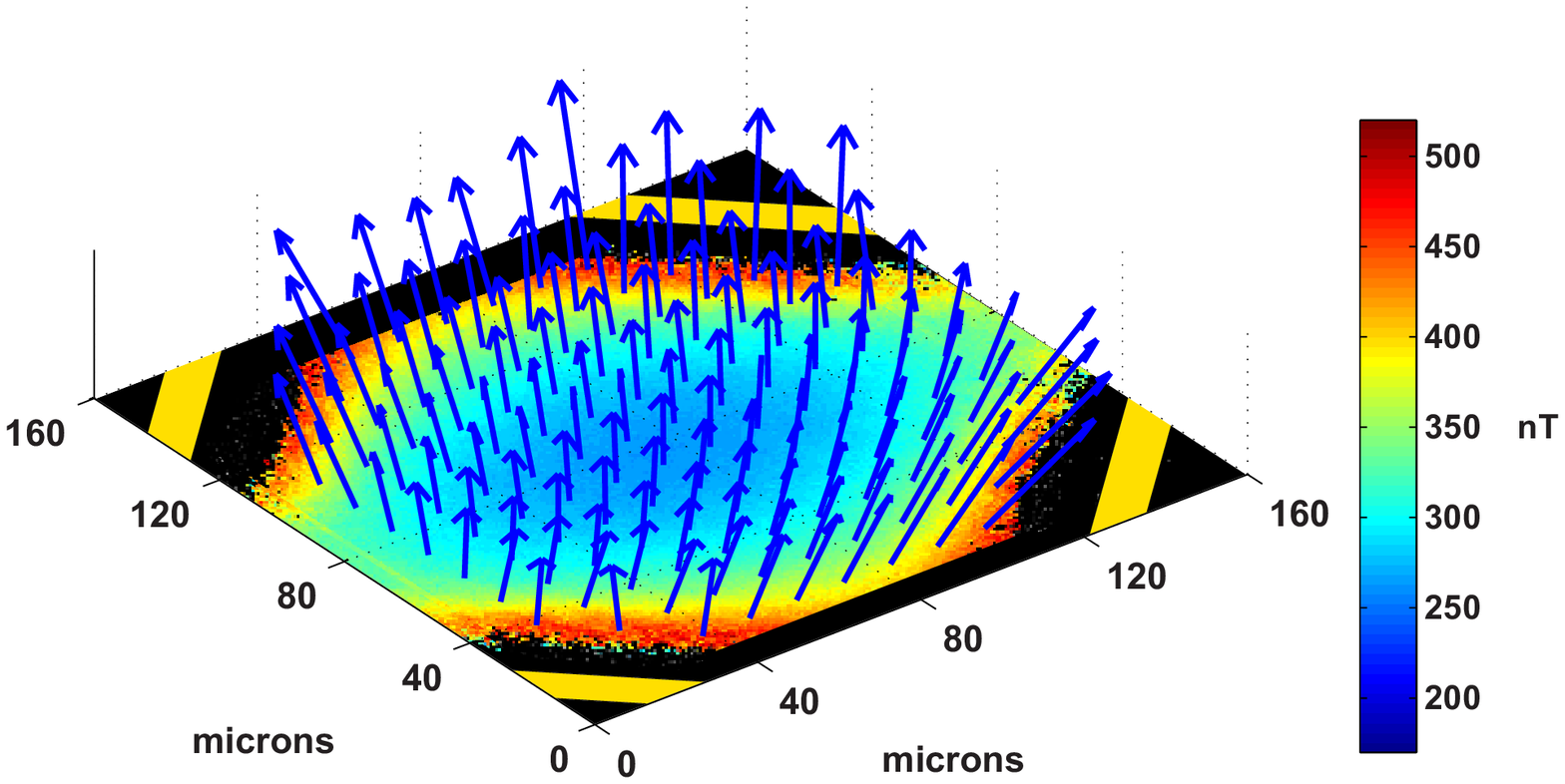}
\label{fig:zigzag:vector}
}
\subfigure[]{
\includegraphics[width=0.4\columnwidth]{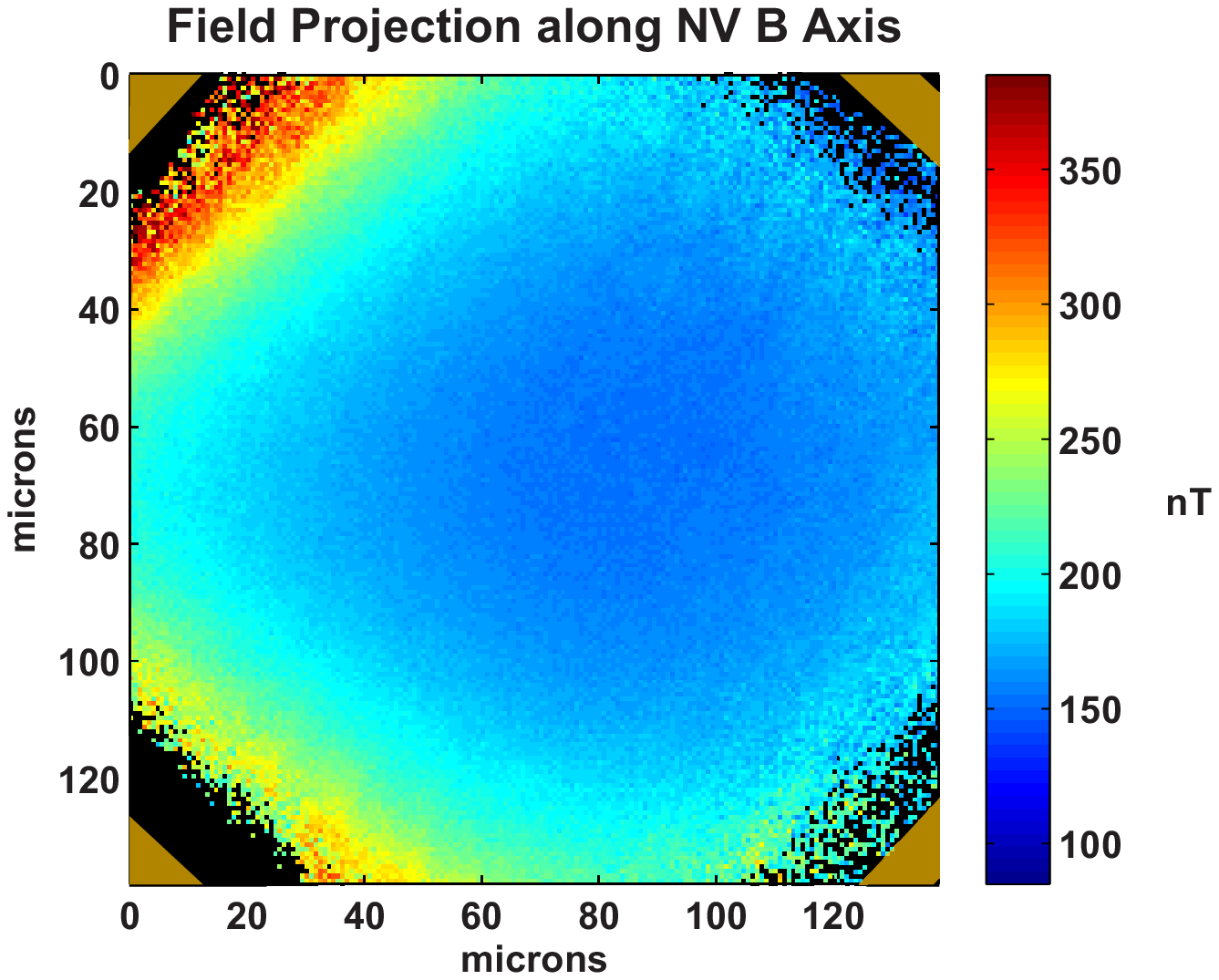}
\label{fig:zigzag:nvb}
}
\label{fig:zigzag}
\caption{~\subref{fig:zigzag:diagram} Diagram of zigzag wire pattern orientation with respect to the four NV axes; map of the field projection along~\subref{fig:zigzag:nva} NV A and~\subref{fig:zigzag:nvb} NV B axes measured when an AC current of frequency 4.75 kHz and amplitude 50 $\mu A$ is run through the zigzag wire pattern; and~\subref{fig:zigzag:vector} a vector field map of the magnetic field reconstructed from field projections measured along two NV axes.}
\end{figure}

The wire pattern was positioned with respect to the diamond chip such that its plane of symmetry coincided with the plane containing the NV axes `NV B' and `NV D' (see Fig.~\ref{fig:zigzag:diagram}). We first used a permanent magnet to apply a static field along the `NV A' axis, resonantly coupled microwaves to that NV population, and ran 4.75 kHz AC current through the wire pattern. Following the same measurement procedure described in the previous section, we extracted a 2D map of the resultant AC magnetic field projection along that axis, as sensed by the ensemble of NV centers near the surface of the diamond (Fig.~\ref{fig:zigzag:nva}). Applying a static field of equal magnitude along the `NV B' axis produced a 2D map of the magnetic field projection along a second axis (Fig.~\ref{fig:zigzag:nvb}). Taking advantage of the two-fold rotational symmetry of the diamond-wire configuration, we were able to use the two maps to reconstruct the magnetic field vector in the plane occupied by the ensemble of NVs (Fig.~\ref{fig:zigzag:vector}).

To determine the sensitivity of the NV magnetic field microscope under the above imaging conditions, we applied a uniform 4.75 kHz AC field across the region of interest and measured the standard deviation of the fluorescence signal at each pixel. Again, the measurements were conducted at an applied field $\rm{B_{AC}}$ with maximum slope in order to optimize the sensitivity. The resulting average-best sensitivity per 614 nm $\times$ 614 nm pixel was $\rm{\approx}$ 136 $\rm{nT/\sqrt{Hz}}$. Substituting experimental parameters such as mean photons collected per measurement, contrast, and measurement time, we calculated an expected shot-noise limited sensitivity per pixel $\rm{\approx}$ 86 $\rm{nT/\sqrt{Hz}}$. The measured sensitivity is once again in reasonable agreement with the shot-noise limited expected value with the discrepancy attributable to technical noise from the laser.

\subsection{Outlook for Improved Sensitivity}
In this work, we demonstrated an NV ensemble magnetic field imager capable of simultaneously measuring spatially varying ac magnetic field patterns over large fields of view ($\rm{\sim 100\:\mu m}$) with sub-micron spatial resolution and achieving echo-based magnetic field sensitivity $\rm{\sim 100\:nT \mu m^{3/2} / \sqrt{Hz}}$~\cite{footnote1}. The preliminary results presented here are an important first step toward potential biological and materials science applications of the NV ensemble magnetic field imager. However, we anticipate that a number of technical advances will be required for this device to reach its full potential. In this section, we outline some possible improvements.

\begin{figure}
\subfigure[]{
\includegraphics[width=0.37\columnwidth]{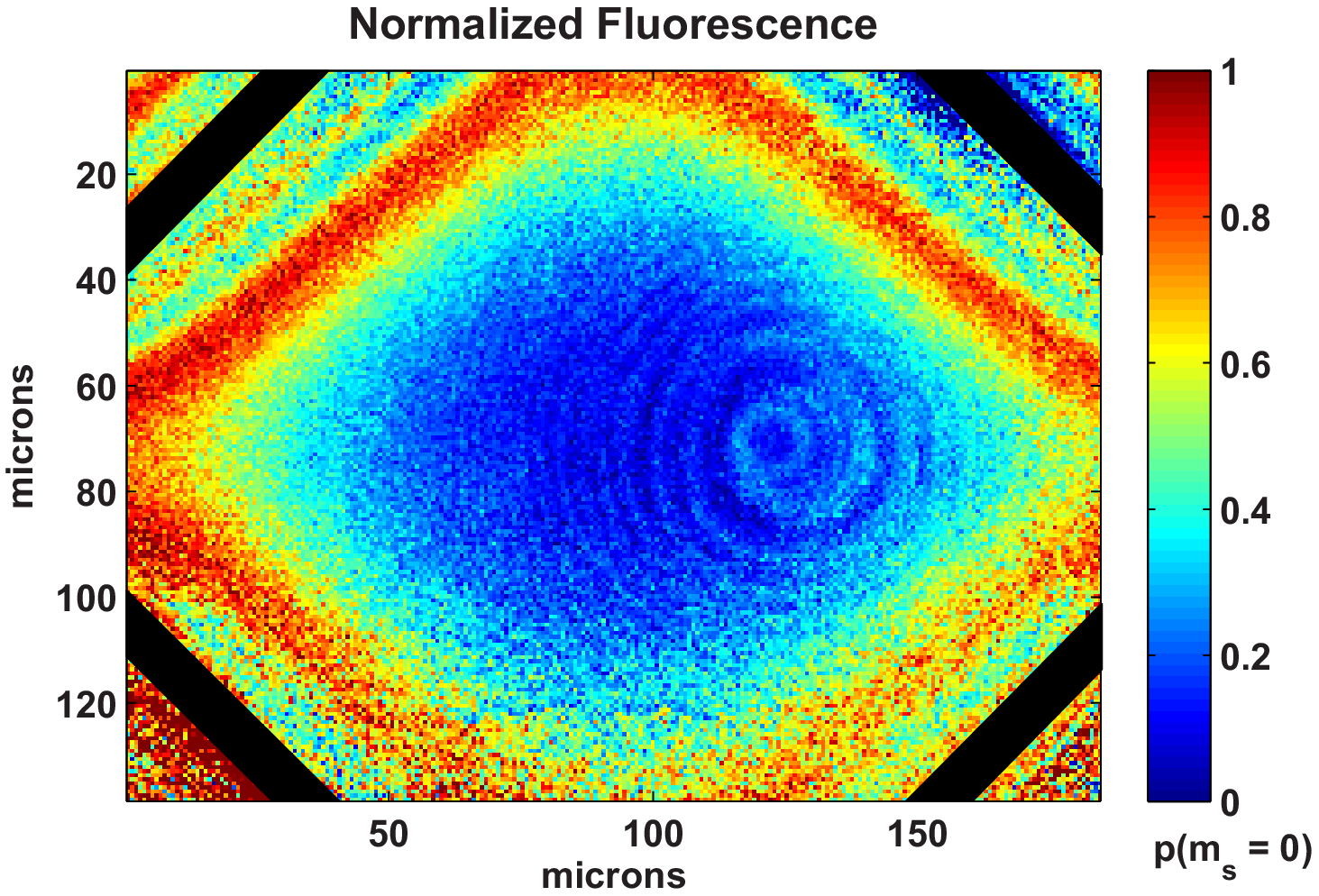}
\label{fig:arc:artifact}
}
\subfigure[]{
\includegraphics[width=0.275\columnwidth]{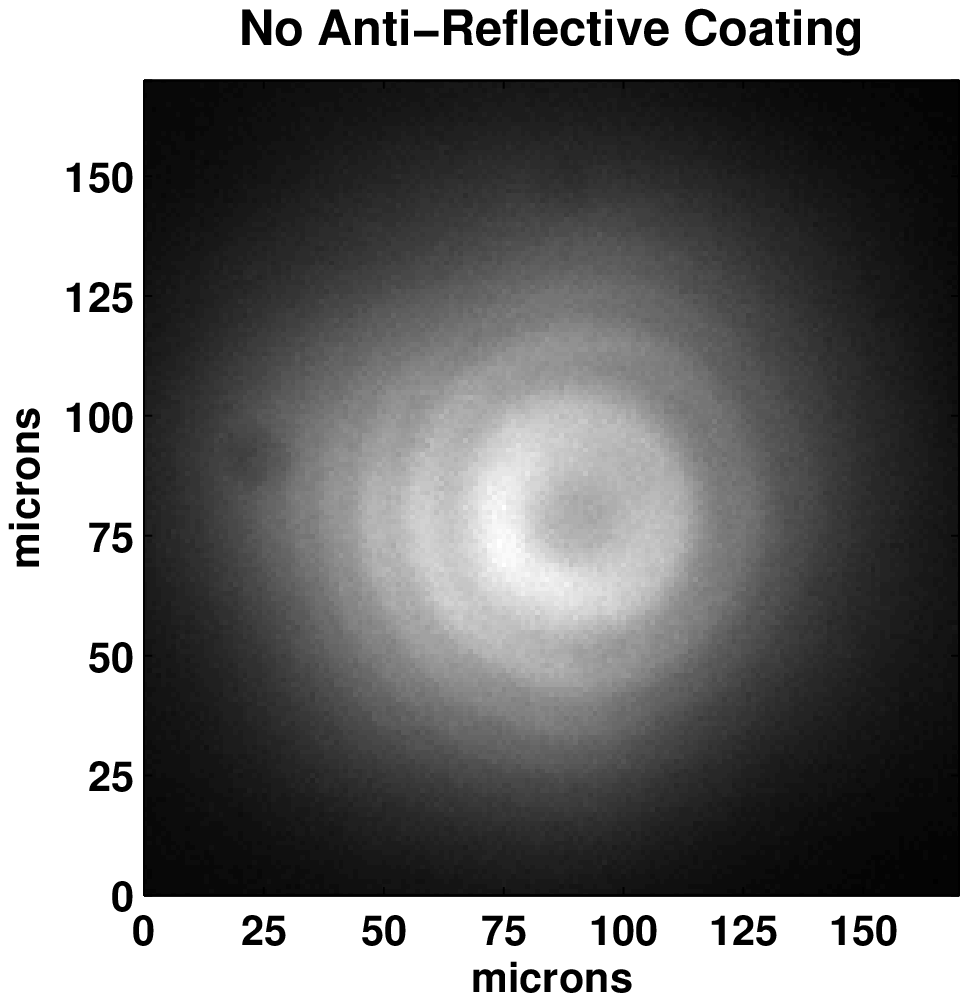}
\label{fig:arc:noar}
}
\subfigure[]{
\includegraphics[width=0.275\columnwidth]{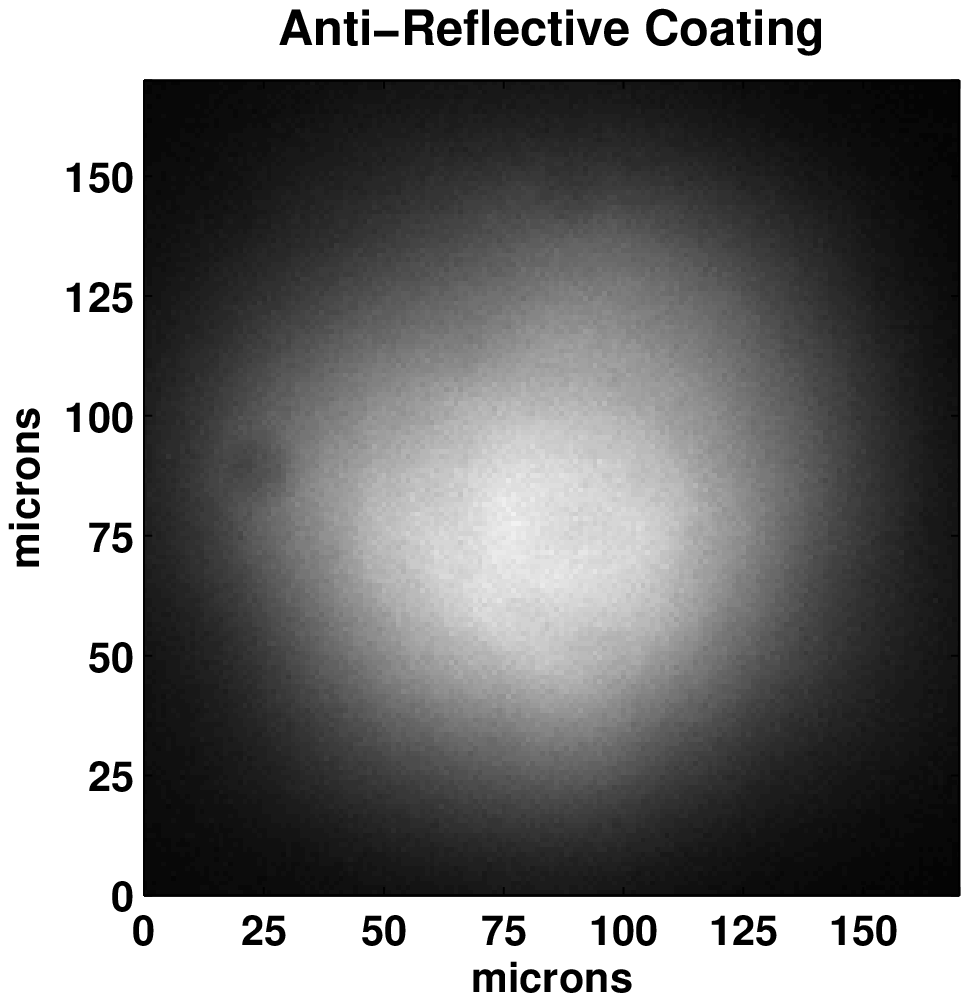}
\label{fig:arc:ar}
}
\label{fig:arc}
\caption{~\subref{fig:arc:artifact} Interference fringe artifacts in NV fluorescence image normalized to display the probability $p$ of the NV population occupying the $m_s = 0$ spin state. This type of raw image is a precursor to an AC magnetic field map.~\subref{fig:arc:noar} NV fluorescence image of a diamond sample without anti-reflective coating. The interference fringes are caused by etaloning effects.~\subref{fig:arc:noar} NV fluorescence image of the same diamond sample with both surfaces anti-reflective coated.}
\end{figure}

One improvement currently being implemented is anti-reflective (AR) coating of the diamond surfaces. The high index of refraction of diamond (n = 2.4) produces $\rm{\approx 17\%}$ reflection of light at the air-diamond interface. Laser illumination of a large area on the diamond produces interference patterns in the incident green and fluoresced red light intensity profiles due to multiple reflections within the diamond (see Fig.~\ref{fig:arc:noar}). These intensity fluctuations produce spatial variations in the amplitude and contrast of the fluorescence signal, resulting in artifacts in the measured magnetic field image (see Fig.~\ref{fig:arc:artifact}). We demonstrated that AR coating via deposition of a quartz layer with $\rm{\lambda/4}$ optical thickness---on either or both surfaces of the diamond chip---drastically reduces reflection-induced artifacts (see Fig.~\ref{fig:arc:ar}). Furthermore, AR coatings reduce reflection of incident excitation light and outgoing fluorescence light at the air-diamond interface, improving the collection efficiency and sensitivity of the measurements.

In these demonstrations, AC magnetic fields were measured along one of four possible crystallographic NV orientations at a given time. NV centers with other crystallographic orientations in the field of view did not contribute to the magnetic field measurement, but did add to the total fluorescence background. Although the excitation light polarization was chosen to minimize the coupling to these other NV orientations, the net result of this effect was a $\sim 3$-fold reduction of the state dependent fluorescence contrast compared to that seen in single-NV confocal experiments. As such, better sensitivity could be achieved by either engineering diamonds with a preferred NV orientation, decreasing the fluorescence signal from undesired NV orientations, or by combining simultaneous magnetometry measurements along the four vectors parallel or antiparallel to the NV orientations to extract the field along their added vectors. The latter option could be implemented in a diamond with reduced concentrations of $\rm{^{13}C}$, in order to ease the stringent requirement of aligning the external magnetic field with a given NV axis for maximal $\rm{T_2}$ coherence time, and would result in a factor of $\sim 2$ improved vector magnetometer sensitivity.

One path toward improved AC magnetometry is to extend the NV coherence times through dynamical decoupling techniques using advanced microwave pulse sequences. For single NV centers, these techniques have demonstrated 7-fold improvements in coherence time and improved magnetometry by a factor of 2 in diamonds with similar coherence properties to the one used in our demonstration of wide-field magnetic field imaging~\cite{RyanPRL2010, LangeSci2010, NaydenovPRB2011}. Therefore, we expect these techniques to be directly applicable to NV ensembles and magnetic field imaging in future experiments.

Further improvements are expected through proper engineering of diamond samples. Longer coherence times should be possible by using diamonds that have reduced concentrations of the $\rm{^{13}C}$ isotope~\cite{BalasubramanianNatMat2009}. An additional magnetometry improvement could be gained by increasing the N-to-NV conversion efficiency in diamonds. The diamond chips used in this work have estimated conversion efficiencies of 3\% for the CVD-grown sample and 8\% for N ion implanted sample, although the expected upper limit is 50\%, assuming one electron-donating N per negatively charged NV center. Greatly enhanced conversion efficiencies have been demonstrated by introducing vacancies into the diamond with ion or electron beams, or high-energy neutron irradiation, followed by a high-temperature annealing process~\cite{MitaPRB1996, MartinAPL1999, PezzagnaNJP2010, NaydenovAPL2010}. We are actively investigating these improvements.

To form a realistic estimate for the magnetic field sensitivity of the optimized NV ensemble magnetic field imager, we assume a diamond sample with similar N concentration as in Section 3.2 but reduced $\rm{^{13}C}$ concentration, resulting in a factor of $\sim 2$ longer coherence time. We further assume a 1 $\rm{\mu m}$ thick NV-enhanced layer for improved magnetic field spatial resolution, and a 30\% N-to-NV conversion efficiency, resulting in a three-fold improvement in the fluorescence signal. Additional planned technical upgrades include improving the fluorescence collection efficiency by switching to an oil-immersion objective with a higher NA, increased optical excitation intensity, laser intensity stabilization and an improved method of monitoring long-term laser power drift, increased microwave field magnitude and homogeneity, AR coating of the sample, magnetometry using all NV orientations, and dynamical decoupling techniques to extend the coherence time. Considered together, these improvements yield a projected sensitivity $\rm{\approx}$ 1 $\rm{nT/\sqrt{Hz}}$ for a single 1 $\rm{\mu m ^{3}}$ pixel. New approaches for improved optical collection efficiency, such as exploiting the diamond substrate as a lens~\cite{HaddenAPL2010, SiyushevAPL2010}, or using the refractive index mismatch between diamond and air to guide light~\cite{BabinecNatNano2010}, may allow NV ensemble magnetic field imaging with sensitivity $\rm{< 1 nT/\sqrt{Hz}}$ per 1 $\rm{\mu m ^{3}}$ pixel.

\section{Application to Imaging of Neuronal Network Dynamics}
With the improvements outlined above, we plan to apply the NV ensemble magnetic field imager to studies of connectivity and signaling in functional, cultured networks of neurons grown on a diamond magnetometer chip, via real-time imaging of the patterns of magnetic fields produced by firing neurons in the network (i.e., action potential propagation down the neurons' axons). Neuronal networks are collections of neurons interconnected by synapses and provide the physical basis of central and peripheral nervous systems. Despite rapid advances in neuroscience, a clear understanding has yet to emerge of how microscopic connectivity of neurons encodes macroscopic function of the network. Determining the rules for translating neuronal connectivity to network function will require experimental tools that are capable of real-time mapping of both local and global spatiotemporal connectivity among neurons and signal propagation between them.

However, given the size of a biologically relevant, functional neuronal network  ($\sim 1$ mm or larger), such real-time functional imaging is a formidable challenge beyond the capabilities of existing experimental instruments and methodologies. An optimized NV ensemble magnetic field imager may be able to address this challenge by providing micron-scale mapping of magnetic fields on a sub-millisecond time scale for all neurons within a functional, cultured network in a minimally invasive manner, following stimulation and/or reward administration, to identify the presence and emergence of persistent patterns of local and global activity. For a separation of $\sim 1$ $\rm{\mu m}$ between the neurons and the NV ensemble, the typical magnetic field pulse generated by a single firing axon occurs on a timescale $\approx 1$ ms, with a peak magnitude that is expected to vary from $\sim 1-100$ nT, depending on species and type of neuron~\cite{WikswoClinNeuro1991}. In early applications of the NV ensemble magnetic field imager, we plan to sense the time-averaged magnetic field from repeated, induced firing of a single large neuron or neuron bundle; e.g., from a frog sciatic nerve (magnetic field $\sim 10$ nT at a distance of 1 $\rm{\mu m}$ from the axon).  As the sensitivity of the imager is improved, we will proceed to study single-shot firing of a single neuron, and then mapping of the real-time electromagnetic dynamics of all neurons within a complex, functional network. From such data, basic parameters for modeling the neuronal network's behavior will be extracted and then compared with leading theoretical models. The merits of each model will be determined by how well it predicts the network's evolution toward the final connectivity state, given the initial state, geometry, and incident inputs as constraints.

\section{Acknowledgments}
This work was supported by NIST, NSF, and DARPA. We gratefully acknowledge the provision of diamond samples by Apollo Diamond, Inc. and technical discussions with Patrick Doering, Jeronimo Maze, and Daniyar Nurgaliev.

\section*{References}
\bibliographystyle{unsrt}
\bibliography{njpbib}

\end{document}